\documentclass[physrev,aps,pra,twocolumn, a4paper,longbibliography,groupedaddress,nofootinbib]{revtex4-2}

\pdfoutput=1

\newcommand{\bfvec}[1]{\mathbf{#1}}
\newcommand{\bfunitvec}[1]{\hat{\mathbf{#1}}}

\newcommand{\e}[1]{\mathrm{e}^{#1}}
\newcommand{\channel}[1]{\{#1\}}

\newcommand{\switch}[1]{\hat{#1}}

\usepackage{float}

\usepackage{fourier}

\usepackage{color}
\usepackage{graphicx}
\usepackage{amsmath,amssymb}

\usepackage{times}
\usepackage{upgreek}
\usepackage{psfrag} 
\usepackage{latexsym} 
\usepackage{amstext}
\usepackage{amsxtra} 

\usepackage{textcomp}
\usepackage{amsfonts}
\usepackage{graphicx}
\usepackage{bm}
\usepackage{color}
\usepackage{braket}
\usepackage[svgnames]{xcolor}

\usepackage{gensymb}
\usepackage{bbold}
\usepackage{siunitx}
\usepackage[normalem]{ulem}

\definecolor{myColor}{rgb}{0.02,0.12,0.3}
\definecolor{myciteColor}{rgb}{0.39,0.7,0.89}
\usepackage[colorlinks=true,citecolor=myColor,linkcolor=myColor,urlcolor=myColor]{hyperref}

\begin{document} 

\title{
Systematic errors arising from polarization imperfections in measurements of the electron's electric dipole moment
}

\author{C. J. Ho}
\thanks{These authors contributed equally to this work.}
\altaffiliation{Present address: Cavendish Laboratory, University of Cambridge, J. J. Thomson Avenue, Cambridge CB3 0HE, United Kingdom}

\author{S. C. Wright}
\thanks{These authors contributed equally to this work.}
\altaffiliation{Present address: Fritz-Haber-Institut der Max-Planck-Gesellschaft, Faradayweg 4-6, 14195 Berlin, Germany}

\author{B. E. Sauer}
\author{M. R. Tarbutt}

\affiliation{Centre for Cold Matter, Blackett Laboratory, Imperial College London, Prince Consort Road, London SW7 2AZ, UK}

\begin{abstract}
The electron's electric dipole moment (eEDM) can be determined by polarizing the spin of an atom or a molecule and then measuring the spin precession frequency in an applied electric field. Radiation is used to polarize the spin and then analyze the precession angle, and the measurement is often sensitive to the polarization of this radiation. We show how systematic errors can arise when both the polarization of the radiation and the magnitude of the electric field are imperfectly controlled. We derive approximate analytical expressions for these errors, confirm their accuracy numerically, and show how they can be corrected empirically. We consider spin manipulation using single-photon pulses, Raman pulses, and Stimulated Raman Adiabatic Passage (STIRAP), and show that STIRAP provides better immunity to these systematic errors. An experimental study of these errors partly supports our findings but also reveals another potential error that is not captured by this analysis.
\end{abstract}

\maketitle

\section{Introduction}

Despite its many successes, the Standard Model is thought to be incomplete, in part because it cannot explain cosmological observations such as dark matter, the matter-antimatter asymmetry, and the accelerating expansion of the Universe. Measurements using atoms and molecules can detect signatures of physics beyond the Standard Model~\cite{Safronova2018}. In particular, experiments that measure the electron's electric dipole moment (eEDM) look for time-reversal-symmetry-violating physics which can be important in resolving the open question of how matter came to dominate the Universe~\cite{Chupp2019}. Heavy atoms such as Cs~\cite{Murthy1989} and Tl~\cite{Regan2002} were used in earlier eEDM experiments as the relativistic motion of the electron near the heavy nucleus enhances the interaction such that the measured atomic EDM can be two orders of magnitude larger than the eEDM~\cite{Schiff1963, Hinds1997}. Heavy polar molecules can provide even greater enhancement because they are more easily polarised in an external electric field, resulting in effective electric fields of $10-100\,\si{GV/cm}$~\cite{Hinds1997}. For over a decade, the most sensitive eEDM measurements have all used molecules, starting with YbF~\cite{Hudson2011}, then ThO~\cite{Baron2014, Andreev2018} and HfF$^+$~\cite{Cairncross2017, Roussy2023}. At present, the most precise upper limit, $|d_e| < \SI{4.1e-30}{e.cm}$, constrains new physics at mass scales above \SI{10}{TeV}~\cite{Roussy2023}. Future experiments aim to improve on this limit by using laser-cooled molecules such as YbF~\cite{Fitch2020b}, BaF~\cite{Aggarwal2018} and YbOH~\cite{Augenbraun2020}, new species of molecular ions such as ThF$^+$~\cite{Zhou2019}, or a large number of molecules trapped in a rare-gas matrix~\cite{Vutha2018}.

A typical eEDM experiment can be described as an atomic or molecular spin precession measurement. The spins of the particles are prepared along an axis perpendicular to the applied electric field $\mathbf{E} = E\bfunitvec{z}$ and allowed to precess freely for a time $\tau$. The precession angle is measured and the eEDM is proportional to that part of the angle that correlates with the direction of $\mathbf{E}$. Systematic effects in these measurements can be conveniently divided into two classes. The first occur during the free evolution time. An example is a magnetic field which changes when $\mathbf{E}$ is reversed. Such effects can be managed by careful control of the static fields in the experiment. In the second class are effects that occur during the preparation and readout of the spin. These require control of light fields which can be challenging due to the interaction of the light with the materials of the apparatus such as the vacuum windows or the electric field plates. It is usually necessary to control the frequency, phase, amplitude and polarization of the light, and a failure to adequately control any one of these can lead to errors. The polarization of the light is often used to prepare and analyze the spin polarization, so the former is critical but can also be the most difficult part of a light field to control. This paper focuses on systematic errors in eEDM measurements arising from polarization imperfections.

\section{Model experiment}\label{sec:modelExperiment}

We consider a simple diatomic molecule such as YbF or BaF, though our analysis can be extended to other molecules with different structures.  We focus on hyperfine levels $F=0,1$ within the rotational ground state, and use the notation $\ket{0} \equiv \ket{F=0,m_F=0}$ and $\ket{\pm1} \equiv \ket{F=1,m_F=\pm1}$. The state $\ket{F=1,m_F=0}$ plays no role because it is Stark-shifted away from $\ket{\pm1}$, whereas $\ket{\pm 1}$ remain degenerate in the electric field. We also define $\ket{x} \equiv i\left(\ket{+1} - \ket{-1}\right)\sqrt{2}$ and $\ket{y} \equiv \left(\ket{+1} + \ket{-1}\right)/\sqrt{2}$. The transition frequency between $\ket{0}$ and $\ket{\pm 1}$ (in the presence of $\mathbf{E}$) is $\omega_0$.

\begin{figure*}
\centerline{\includegraphics[width=2\columnwidth]{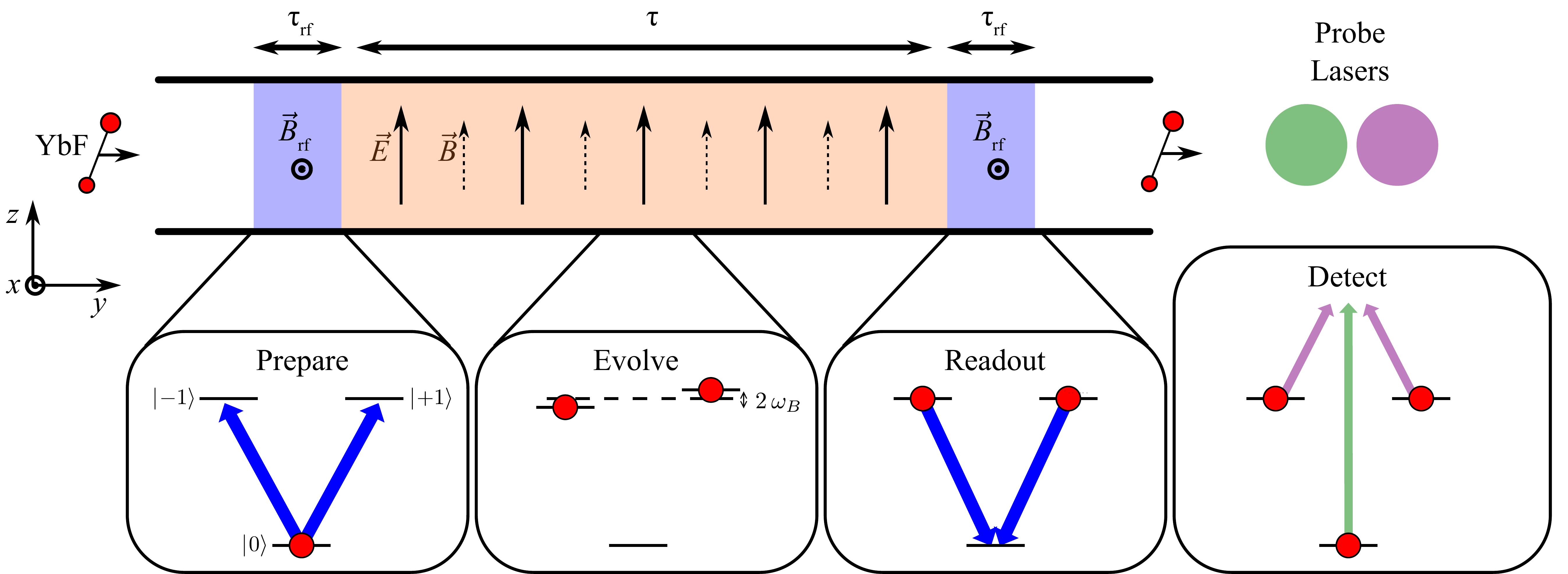}}
\caption{Schematic illustrating an idealised eEDM measurement using a YbF molecular beam, and the coordinate system used throughout this paper. Molecules enter the experiment in the state $\ket{0}$ and are prepared in the state $\ket{x}$ by a pulse of rf radiation polarised along the $x$-axis (left blue shaded region). The molecular spin then precesses in static electric and magnetic fields applied along the $z$ direction for a time $\tau$ (orange shaded region). A second rf pulse (right blue shaded region) converts the precession angle $\varphi$ into a population asymmetry between the $F=0$ and $F=1$ hyperfine levels, and this asymmetry is then measured in the detection region. 
}
\label{fig:IFoverview}
\end{figure*}

The ideal experiment using YbF molecules is illustrated in Fig.~\ref{fig:IFoverview}. Molecules are assumed to be prepared in the state $\ket{0}$, and are then transferred to $\ket{x}$ (or $\ket{y}$) by light polarized along $\bfunitvec{x}$ (or $\bfunitvec{y}$). This can be done using a resonant single-photon (rf) process or a resonant two-photon (optical) process.
After a free evolution time $\tau$ in the electric ($\mathbf{E} = E\bfunitvec{z}$) and magnetic ($\mathbf{B} = B\bfunitvec{z}$) fields, the state $\ket{x}$ evolves into $i\left(\e{-i\varphi}\ket{+1} - \e{i\varphi}\ket{-1}\right)/\sqrt{2} = \cos\varphi\ket{x} + \sin\varphi\ket{y}$, where $\varphi = \left(\mu B - d_e E_\mathrm{eff}\right)\tau/\hbar$ is referred to as the interferometer phase. Here, $\mu$ is the magnetic moment, $d_e$ is the eEDM, and $E_\mathrm{eff}$ is the effective electric field. A second interaction with light of the same polarization results in the state $\cos\varphi\ket{0} + \sin\varphi\ket{y}$. The populations in $F=0$ and $F=1$ are measured, and are proportional to $\cos^2\varphi$ and $\sin^2\varphi$ respectively. Their difference divided by their sum gives the quantity called the asymmetry, $\mathcal{A}$, which in an ideal experiment is $\mathcal{A} = \cos2\varphi$. We write $\varphi = \phi_B + \phi$ where $\phi_B = \mu B \tau/\hbar = \omega_{B}\tau$ is the magnetic part of the phase and $\phi$ is a very small additional phase due to the eEDM or arising from imperfections in the experiment. To maximize the sensitivity to $\phi$, we typically set $B$ such that $\phi_B = \switch{B}\pi/4$, where $\switch{B}=\pm1$ indicates the direction of $\mathbf{B}$. In this case, $\mathcal{A} = -2 \switch{B}\phi $.

One powerful way to diagnose and avoid potential systematic errors is to modulate the important parameters between shots of the experiment. In our experiment, we reverse the directions of $\mathbf{E}$ and $\mathbf{B}$ and step the amplitude and frequency of the light around their ideal values. We also introduce a $\pm \pi/2$ phase shift between the two light fields so that any Ramsey-type signal due to coherence between the residual population in the $F=0$ state and the population in $F=1$ is averaged to zero.
Each switch $X$ has two possible states which we write as $\switch{X} = \pm 1$. The asymmetry values correlated with each of these switches (or a combination of these switches), which we call channels, provide valuable information about the experiment. The asymmetry that correlates with the product of  $\switch{X}_1,\switch{X}_2,\ldots,\switch{X}_m$ is 
\begin{equation}
    \channel{X_1\cdot X_2\cdot\ldots\cdot X_m} = \frac{1}{N} \sum_{i=1}^N \prod_{j=1}^m \switch{X}_{j,i}\mathcal{A}_i,
\label{eq:channelDefn}
\end{equation}
where the subscript $i$ denotes the $i^\mathrm{th}$ shot of the experiment, and $\switch{X}_{j,i}$ is the state of parameter $X_j$ during shot $i$. For example, the asymmetry that correlates with the frequency step $\delta$ of the state preparation or readout field, $\channel{\delta}$, is proportional to the value of the mean detuning from resonance and can be used to minimize long term drifts in the detuning. Similarly, the asymmetry that correlates with the direction of $\mathbf{B}$, $\channel{B}$, gives the background magnetic field and can be used to ensure we operate at zero field. The asymmetry that correlates with the directions of both $\mathbf{E}$ and $\mathbf{B}$, $\channel{E\cdot B}$, gives the interferometer phase correlated with the direction of $E$ --- the eEDM-induced phase appears in this channel.

We investigate a systematic effect arising from two parts. The first part, which is our main focus, arises when the light field used for state preparation and readout has some ellipticity. This effect leads to a non-zero $\channel{B \cdot \delta}$ value, and so has the signature of an interferometer phase that depends on the detuning of the light field.
The second part arises when the magnitude of $\mathbf{E}$ changes upon reversal. This changes the resonance frequency of the $\ket{0} \rightarrow \ket{x}$ transition due to the Stark shift, leading to a non-zero $\channel{E\cdot \delta}$ value, which can be interpreted as a detuning of the light correlated with the direction of $\mathbf{E}$. The combination of these two effects results in an interferometer phase correlating with the direction of $\mathbf{E}$, which is the same signature as the eEDM. This effect has been observed in an eEDM measurement using YbF~\cite{Hudson2011, Kara2012}, which is the basis for this paper. Similar effects have also been observed in the ThO eEDM experiments~\cite{Baron2014, Andreev2018}.

\section{Single-photon rf pulses}
\label{sec:rfPulses}
We first consider a single-photon process for state preparation and readout. Typically, this is done with two rf $\pi$-pulses, each polarised along $\bfunitvec{x}$, which transfer population between $\ket{0}$ and $\ket{x}$. We introduce imperfections to the rf polarization with two transformation matrices, $P(\epsilon)$ and $R(\theta_{\mathrm{rf}})$, given by,
\begin{equation}
P(\epsilon) = \frac{1}{\sqrt{1+\epsilon^2}}
    \begin{pmatrix}
        1 & -i\epsilon \\
        i\epsilon & 1 \\
    \end{pmatrix}
    \hspace{0.1cm}, \hspace{0.1cm}
    R(\theta_{\mathrm{rf}}) = 
    \begin{pmatrix}
        \cos\theta_{\mathrm{rf}} & -\sin\theta_{\mathrm{rf}} \\
        \sin\theta_{\mathrm{rf}} & \cos\theta_{\mathrm{rf}}\\
    \end{pmatrix}
    \hspace{0.1cm}.
\end{equation}

\noindent When acting on a linearly polarised field, $P(\epsilon)$ introduces an ellipticity, with $-1\leq\epsilon\leq 1$ and where the limits correspond to circular polarizations of opposite handedness. $R(\theta_{\mathrm{rf}})$ rotates the axes of the ellipse about the $z$-axis by the angle $\theta_{\mathrm{rf}}$. Since $P(\epsilon)$ and $R(\theta_{\mathrm{rf}})$ commute, the order in which they are applied is unimportant.

\begin{figure}
\centerline{\includegraphics[width=0.9\columnwidth]{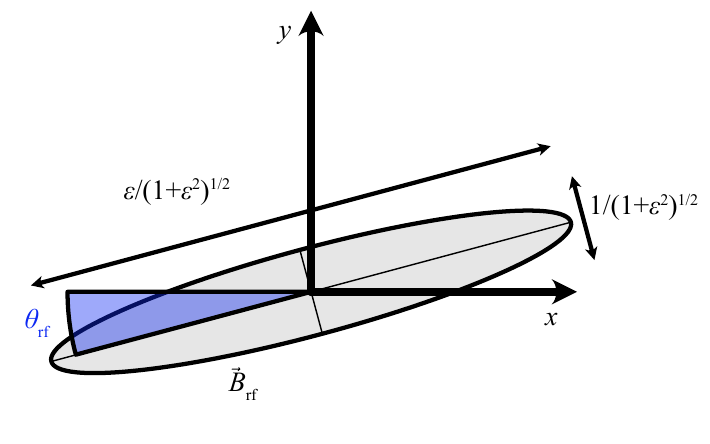}}
\caption{A sketch of the rf magnetic field in the $x$-$y$ plane. The parameters $\epsilon$ and $\theta_{\mathrm{rf}}$ defining the experimental imperfections are discussed in the text.}
\label{fig:polDiagram}
\end{figure}

The imperfect rf polarization in the experiment is expressed as $\bfvec{B}_{\mathrm{rf}} = B_{\mathrm{rf}} \cos(\omega t + \alpha) \bfunitvec{e}$, with $\bfunitvec{e} = P(\epsilon)R(\theta_{\mathrm{rf}})\bfunitvec{x}$. We plot the resulting polarization ellipse in Fig.~\ref{fig:polDiagram}. This rf field, rather than coupling $\ket{0}$ to $\ket{x}$, couples $\ket{0}$ to $\ket{x'}$, which is
\begin{equation}
\begin{aligned}
    \ket{x'} = & \frac{1}{\sqrt{1+\epsilon^2}}\Big[(\cos\theta_{\mathrm{rf}} -i\epsilon\sin\theta_{\mathrm{rf}})\ket{x}\\
&+(i\epsilon\cos\theta_{\mathrm{rf}} + \sin\theta_{\mathrm{rf}})\ket{y}\Big] \\
    =&  \frac{i}{\sqrt{2(1+\epsilon^2)}}\Big[e^{-i\theta_{\mathrm{rf}}}(1+\epsilon)\ket{-1}+ e^{i\theta_{\mathrm{rf}}}(-1+\epsilon)\ket{+1}\Big]
    \label{eq:xprimestate}
\end{aligned}
\end{equation}

\noindent We associate $\epsilon$ with a population imbalance between the $\ket{\pm 1}$ states, and the angle $\theta_{\mathrm{rf}}$ with a relative phase between them. The latter imperfection has the same effect as a background magnetic field, so we neglect this for now.

We allow different frequencies and ellipticities for the two rf pulses and label these with an index $k=1,2$. In the basis $\{ \ket{0}, \e{i\omega_k t}\ket{-1}, \e{i\omega_k t}\ket{+1} \}$, and in the rotating-wave approximation, the Hamiltonian describing the interaction of the molecule with $\mathbf{E}$, $\mathbf{B}$ and $\bfvec{B}_{\mathrm{rf}}$ is
\begin{equation}
    \mathcal{H}_{\mathrm{rf},k} = \hbar
    \begin{pmatrix}
        0 & \frac{\Omega(1-\epsilon_k)e^{i\alpha}}{2\sqrt{2}} & \frac{\Omega(1+\epsilon_k)e^{i\alpha}}{2\sqrt{2}} \\
        \frac{\Omega(1-\epsilon_k)e^{-i\alpha}}{2\sqrt{2}} & -\delta_k-\omega_B & 0 \\
        \frac{\Omega(1+\epsilon_k)e^{-i\alpha}}{2\sqrt{2}} & 0 & -\delta_k+\omega_B
    \end{pmatrix},
\label{eq:hrf}
\end{equation}
where $\Omega = - \bra{\pm 1}\mu_{\mp 1}\ket{0}B_{\mathrm{rf}}/\hbar$ is the Rabi frequency, $\mu_{p}$ are the spherical components of the magnetic moment operator, $\delta_k = \omega_k - \omega_0$ is the detuning, and $\hbar\omega_B = \mu B$ is the Zeeman shift of the $\ket{\pm1}$ states. The Hamiltonian describing the evolution in the static fields between the the two rf pulses, $\mathcal{H}_\mathrm{static}$, is the same as $\mathcal{H}_{\mathrm{rf},1}$ with $\Omega=0$~\footnote{Here, we have used the dressed-state basis defined for $\omega_1$.}. 

The final state of a molecule that starts in $\ket{0}$ is given by 
\begin{equation}
    \ket{\psi_f} = U_\mathrm{rf,2}(\tau_\mathrm{rf},\switch{\pi}\pi/2)U_\mathrm{static}(\tau,0)U_\mathrm{rf,1}(\tau_\mathrm{rf},0)\ket{0},
\label{eq:timeEvolution}
\end{equation}
where $U_{m}(t,\alpha) = \e{-i\mathcal{H}_mt/\hbar}$ is the propagator. The probability of measuring the molecule in $F=0$ is $p_0 = \lvert\braket{0|\psi_f}\rvert^2$, and the asymmetry is $\mathcal{A} = 2p_0 - 1$. 
We introduce four switches: $\switch{B}$, which reverses the direction of the $\mathbf{B}$ field, $\switch{\delta}_1$ and $\switch{\delta}_2$, which changes the sign of a small, intentional detuning of magnitude $\delta$ applied to pulse $k$, and $\switch{\pi}$, which changes the phase of the second rf pulse between $\pm \pi/2$. This last switch removes the effects of unwanted coherences between the $\ket{0}$ and $\ket{\pm1}$ states.  For every combination of switch states $( \switch{B}, \switch{\delta_1}, \switch{\delta_2}, \switch{\pi} )$, we calculate  $\braket{0|\psi_f}$ and expand it to lowest order in the small quantities $\epsilon_1$, $\epsilon_2$, $\delta$ and $\omega_B$. From this, we calculate the asymmetry values and finally the channel values using Eq.~(\ref{eq:channelDefn}). 

\subsection{Simple example -- only the second pulse is imperfect}

\begin{figure}[t!]
\centerline{\includegraphics[width=1\columnwidth]{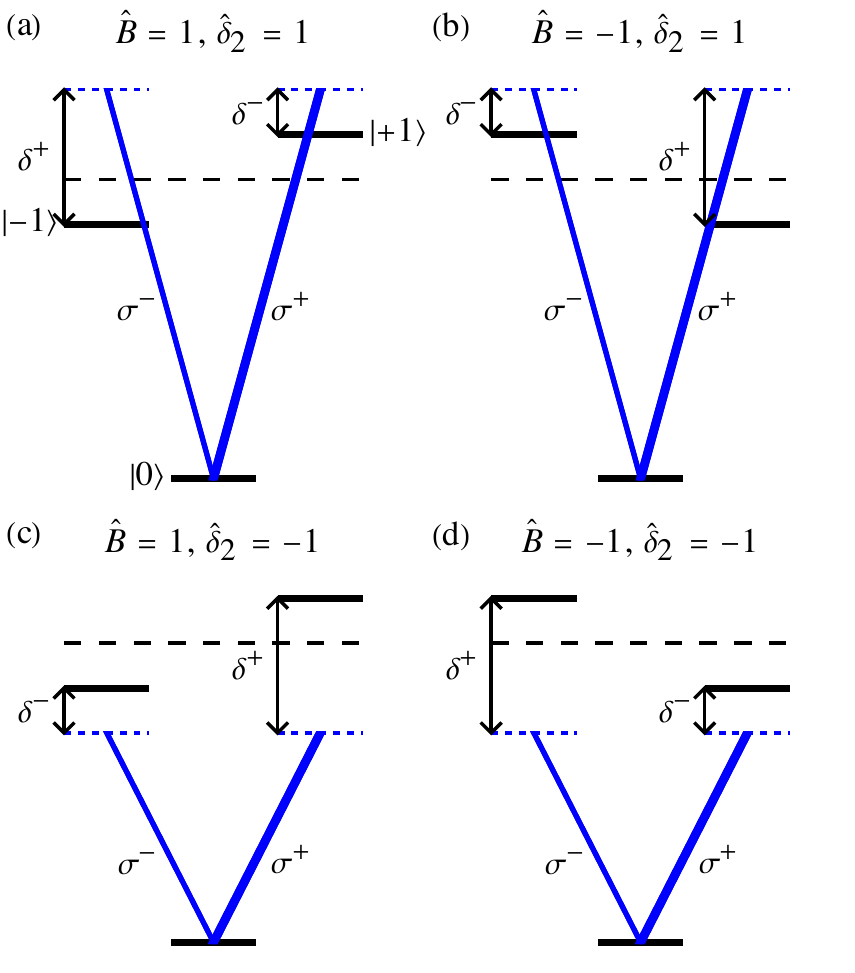}}
\caption{
Origin of the asymmetry correlated with the switches $\switch{B}$ and $\switch{\delta}_2$. Panels (a) -- (d) show the relevant rf transitions (blue lines) in the state detection step of the experiment for different switch values of $\switch{B}$ and $\switch{\delta}_2$. The thicknesses of the lines are indicative of the strengths of the rf transitions; here we illustrate the case where $\epsilon > 0$. The detuning of each transition is given, where $\delta^\pm = \lvert \delta \pm \omega_B \rvert$, $\delta$ is the applied detuning of the rf pulse and $\hbar \omega_B$ is the Zeeman shift.
}
\label{fig1}
\end{figure}

As a simple example, we consider the situation where the first pulse has no polarization imperfection or detuning step, $\delta_1 = 0, \epsilon_1 = 0$, while the second pulse has non-zero ellipticity $\epsilon_2$ and detuning step $\delta$. Panels (a)-(d) in Fig.~\ref{fig1} illustrate the energy levels and rf transitions for the four different combinations of $\switch{B}$ and $\switch{\delta}_2$ in this simple experiment. There is equal population in the $\ket{\pm1}$ states immediately before the second rf pulse. The population transferred to $\ket{0}$ depends on the Rabi frequency and the magnitude of the detuning. In (a), the $\sigma^+$ transition has amplitude $\Omega(1+\epsilon)$ and the detuning $\delta^- \equiv \lvert \delta -\omega_B \rvert$, whereas the $\sigma^-$ transition has amplitude $\Omega(1-\epsilon)$ and the detuning $\delta^+ \equiv \lvert \delta+\omega_B \rvert$. Here, the stronger (weaker) transition has the smaller (larger) detuning. The situation is not reversed symmetrically in (b), where the stronger transition now has the larger detuning, and vice versa. This leads to a difference in population transferred to $\ket{0}$ and a change in asymmetry between (a) and (b). The cases in (c) and (d) are the same as (b) and (a), respectively. With reference to these pictures, the value of the $B\cdot\delta_2$ channel is
\begin{equation}
    \channel{B\cdot\delta_2} = \frac{1}{4}\left(\mathcal{A}_\mathrm{(a)} - \mathcal{A}_\mathrm{(b)} - \mathcal{A}_\mathrm{(c)} + \mathcal{A}_\mathrm{(d)}\right),\nonumber
\end{equation}
which is non-zero when $\epsilon \neq 0$ because of this difference in population transfer from the $\ket{\pm1}$ states. When $\epsilon = 0$ both $\sigma^\pm$ transitions have the same amplitude, and the asymmetry is the same in all four cases. 

Following the procedure outlined in Sec.~\ref{sec:modelExperiment} to calculate the channel values, we find
\begin{equation}
    \channel{B\cdot\delta_2} = \frac{4\epsilon_2\omega_B \delta}{\Omega^2},
\label{eq:brf2fChannelValue}
\end{equation}
which is to lowest order linear in all three small quantities $\epsilon_2$, $\delta/\Omega$ and $\omega_B/\Omega$. Using Eq.~(\ref{eq:channelDefn}) together with $\mathcal{A} = -2 \switch{B}\phi $, we can write 
\begin{equation}
    \channel{B\cdot\delta_2} = \frac{1}{N}\sum_{i=1}^N -2\switch{\delta}_{2,i}\phi_i = -2\phi_{\delta_2},
\label{eq:brffToPhase}
\end{equation}
showing that, despite its origin as an imbalance of population transfer, $\channel{B\cdot\delta_2}$ can be interpreted as a phase that correlates with the detuning of the second rf pulse which we write as $\phi_{\delta_2}$. This interpretation is useful because its value,
\begin{equation}
    \phi_{\delta_2} = -\frac{2\epsilon_2\omega_B \delta}{\Omega^2},\nonumber
\end{equation}
can easily be compared to other phases in the experiment (e.g. the one due to the eEDM).

\begin{figure}[t!]
\centerline{\includegraphics[width=1\columnwidth]{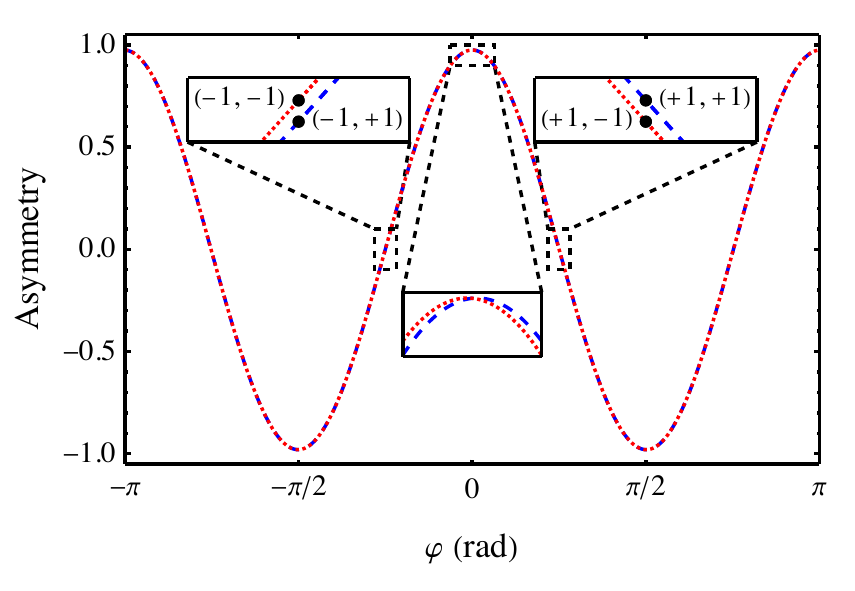}}
\caption{
Asymmetry plotted against the phase produced by the $B$ field, with $\epsilon_1 = 0$, $\epsilon_2 = 0.1$ and $\Omega = \SI{0.1}{MHz}$. The blue (dashed) and red (dotted) curves have $\delta = \pm\SI{5}{kHz}$ respectively. Black filled circles show the values of the switches $(\switch{B}, \switch{\delta}_2)$.
}
\label{fig2}
\end{figure}

Figure~\ref{fig2} shows a numerical calculation of the asymmetry as a function of magnetic field for this experiment, where we have set $\epsilon_2 = 0.1$ and $\delta=\pm\SI{5}{kHz}$, which are realistic parameters for a molecular beam experiment. The insets show the values of the asymmetry at the four switch states. We see that when there is ellipticity, changing the sign of the rf detuning shifts the interference curve, mimicking an interferometer phase shift. 

This effect leads to a false eEDM when combined with an $\mathbf{E}$-correlated rf detuning, $\delta_E$. This arises if the magnitude of $\mathbf{E}$ changes when $\mathbf{E}$ reverses, since this changes the Stark shift of the rf transition. We can include this in our model by adding the $E$ switch to the model and including a term $\switch{E} \delta_E$ in the rf detuning. This immediately leads to a phase which correlates with $E$:
\begin{equation}
    \phi_E = -\frac{2\epsilon_2\omega_B\delta_E}{\Omega^2}.
\label{eq:edmPhaseSimple}
\end{equation}
Fortunately, $\delta_E$ also appears in the channel which gives the asymmetry correlated with $\switch{E}$ and $\switch{\delta}_2$,
\begin{equation}
    \channel{E\cdot\delta_2} = -\frac{2\delta_E\delta}{\Omega^2}.
    \label{eq:Erf2f}
\end{equation}
It follows that the systematic error can be corrected using the measured values of $\channel{B\cdot\delta_2}$ and $\channel{E\cdot\delta_2}$. Using Eqs.~(\ref{eq:brf2fChannelValue}) -- (\ref{eq:Erf2f}), we see that the required correction to the phase is 
\begin{equation}
    \phi_{E,\mathrm{corr}} = -\frac{\Omega^2}{4\delta^2}\channel{B\cdot\delta_2}\channel{E\cdot\delta_2},
\label{eq:edmPhaseCorrSimple}
\end{equation}
where we only need to supply the known parameters $\Omega$ and $\delta$. While this correction removes the systematic error, it will tend to increase the uncertainty of the measurement.

\subsection{Full interferometer}

\begin{figure}[t!]
\centerline{\includegraphics[width=1\columnwidth]{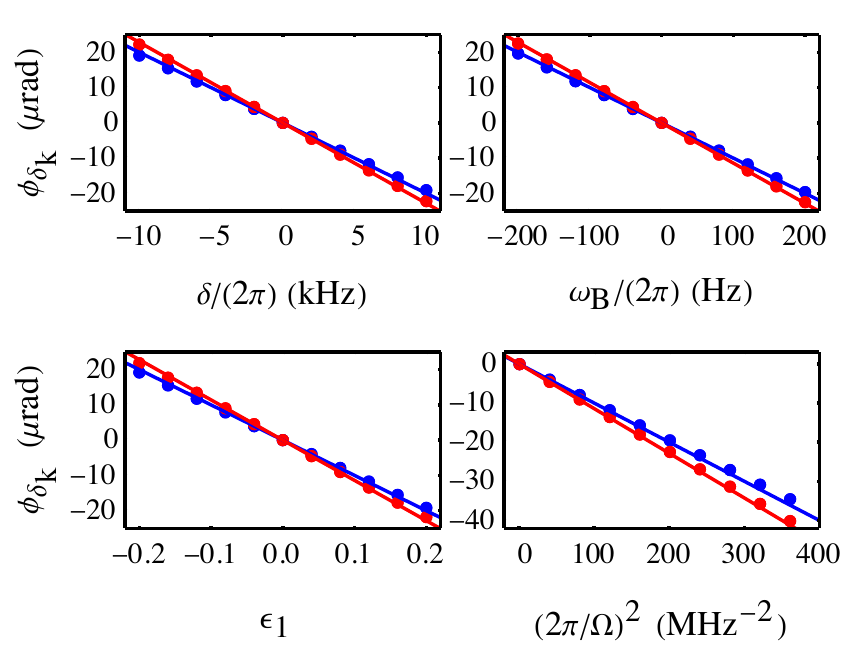}}
\caption{
Comparison of numerical and analytical predictions for the phase correlated with the detunings $\delta_k$ of the rf pulses $k=1$ (blue) and $k=2$ (red). The numerical results are given as filled circles and the analytical expressions, Eqs.~(\ref{eq:brffFullInt}), are plotted as solid lines. The parameters, when not varied, are $\delta/(2\pi) = \SI{5}{kHz}$, $\omega_B/(2\pi) = \SI{100}{Hz}$, $\epsilon_1 = 0.1$ and $\Omega/(2\pi) = \SI{0.1}{MHz}$.
}
\label{fig3}
\end{figure}

Now, we extend the simple example by including the ellipticity and frequency step of both rf pulses. The phases correlated with the detuning of each rf pulse are found to be
\begin{align}
    \phi_{\delta_1} &= -\frac{1}{2}\channel{B\cdot\delta_1} = -\frac{\omega_B\delta}{\Omega^2}\left( 2\epsilon_1 + (2\pi-4)\epsilon_2 \right), \nonumber \\
    \phi_{\delta_2} &= -\frac{1}{2}\channel{B\cdot\delta_2} = -\frac{\omega_B\delta}{\Omega^2}\left( 2\epsilon_2 + (2\pi-4)\epsilon_1 \right).
\label{eq:brffFullInt}
\end{align}
We note that the phase correlating with the detuning of pulse 1 has two terms, one proportional to the  ellipticity of pulse 1 and the other proportional to the ellipticity of pulse 2. The two terms have similar coefficients, since $2\pi-4$ is quite close to 2. The same holds for the phase correlating with the detuning of pulse 2. In order to check these analytical first-order expressions, we compare to numerical results obtained by integrating the time-dependent Schr\"{o}dinger equation for molecules evolving through the interferometer, using the Hamiltonian given by Eq.~(\ref{eq:hrf}). The numerical results are shown in Fig.~\ref{fig3}, where we plot the dependence of $\phi_{\delta_i}$ on $\delta$, $\omega_B$, $\epsilon_1$ and  $\Omega$, and have set $\epsilon_2 = 0$. We see that the numerical results agree well with Eq.~(\ref{eq:brffFullInt}).

As with the simple case considered earlier, the ellipticities lead to a systematic error when combined with an imperfect $\mathbf{E}$-reversal, analogous to Eq.~(\ref{eq:edmPhaseSimple}). Once again, this systematic error can be corrected using other measured channels: 
\begin{equation}
    \phi_{E,\mathrm{corr}} = -\frac{\Omega^2}{4\delta^2}\left(\channel{B\cdot\delta_1}\channel{E\cdot\delta_1}+\channel{B\cdot\delta_2}\channel{E\cdot\delta_2}\right).
\label{eq:edmPhaseCorrFull}
\end{equation}

\subsection{Parameter imperfections}\label{sec:param_imperfections}

There is another effect that causes an apparent rf-detuning-correlated interferometer phase, $\phi_{\delta_i}$, which has nothing to do with elliptical polarizations, but instead is caused by two imperfections. The first is a `background' interferometer phase $\phi_{\rm bg}$ that does not reverse with the switch $\switch{B}$, which arises due to a non-zero background magnetic field, or due to a difference in the polarization angle $\theta_{\rm rf}$ between the two pulses. The second is an rf detuning offset $\Delta_i$ that does not reverse with $\switch{\delta}_i$. These two imperfections separately produce a signal in the channels $\channel{B}$ and $\channel{\delta_i}$, and their combination also results in a signal in $\channel{B\cdot\delta_i}$. Expanding to lowest order in the small quantities $\phi_{\rm bg}$, $\Delta_i/\Omega$ and $\delta_i/\Omega$, we obtain
\begin{align}
    \channel{B} &= -2\phi_\mathrm{bg},\nonumber \\
    \channel{\delta_i} &= -\frac{2 \Delta_i\delta}{\Omega^2},\nonumber \\
    \channel{B\cdot\delta_i} &= -2\phi_{\delta_i} = 4\phi_\mathrm{bg}\frac{\Delta_i\delta}{\Omega^2} = \channel{B}\channel{\delta_i}.\label{eq:offsetBdelta}
\end{align}
 This leads to a potential systematic effect, in the same way as an elliptically-polarised rf field does. However, this effect can be reduced to a negligible value by measuring the channels $\channel{B}$ and $\channel{\delta_i}$ and feeding back to the applied magnetic field and rf frequencies to make them zero. Since all channels are measured with comparable uncertainty, and the effect is proportional to the product of two channels that are both adjusted to zero, the uncertainty in the systematic error is negligible.

\section{Two-photon optical pulses}

\begin{figure}[t!]
\centerline{\includegraphics[width=1\columnwidth]{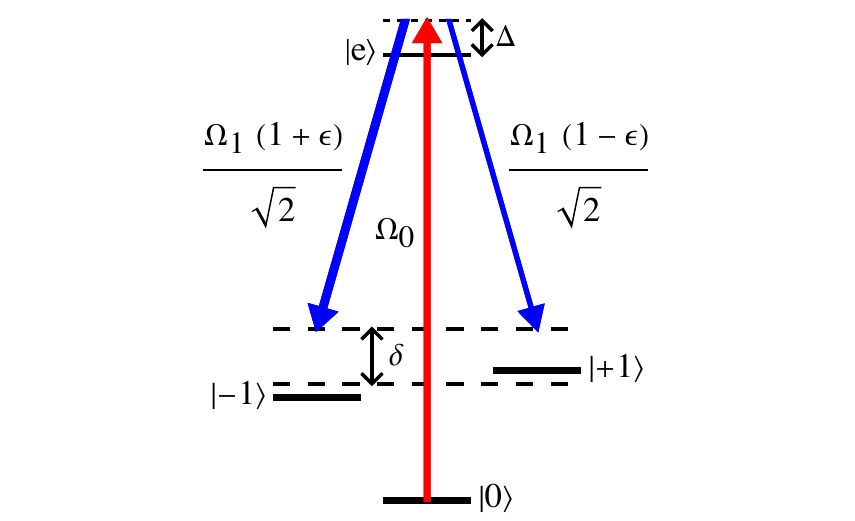}}
\caption{
A two-photon optical transition from $\ket{0}$ to $\ket{x}$. The optical field connecting $\ket{0}$ and $\ket{e}$ (red arrow) is polarised along $\bfunitvec{z}$ and has Rabi frequency $\Omega_0$. A second optical field connects $\ket{e}$ to $\ket{x}$ (blue arrows) and has Rabi frequency $\Omega_1$. Nominally, the latter is polarised along $\bfunitvec{x}$ but some non-zero ellipticity $\epsilon$ might be present, which changes the relative amplitudes of the transitions to $\ket{\pm 1}$.
}
\label{fig4}
\end{figure}

Another way to connect the states of interest is to use a two-photon optical process via an intermediate state, $\ket{e}$, which we take to be an $m_F = 0$ state. One pulse at frequency $\omega_0$ couples $\ket{0} \leftrightarrow \ket{e}$ and is polarised along $\bfunitvec{z}$, while the other at frequency $\omega_1$ couples $\ket{\pm 1} \leftrightarrow \ket{e}$ and is nominally polarised along $\bfunitvec{x}$, but may have some ellipticity $\epsilon$, defined in a similar way to the rf pulses. The energy levels and optical transitions are shown in Fig.~\ref{fig4}, where we have defined $\Omega_0$ and $\Omega_1$ as the Rabi frequencies of the two optical pulses, $\omega_B$ as the Zeeman shift due to a magnetic field, $\Delta$ as the one-photon detuning and $\delta$ as the two-photon detuning. 

The Hamiltonian describing the system illustrated in Fig.~\ref{fig4} is
\begin{equation}
    \mathcal{H}_{\mathrm{Raman}} = \hbar \begin{pmatrix}
        0 & 0 & 0 & \frac{\Omega_0}{2} \\
        0 & -\delta - \omega_B & 0 & -\frac{\Omega_1(1+\epsilon)}{2\sqrt{2}} \\
        0 & 0 & -\delta+\omega_B & \frac{\Omega_1(1-\epsilon)}{2\sqrt{2}} \\
        \frac{\Omega_0}{2} & -\frac{\Omega_1(1+\epsilon)}{2\sqrt{2}} & \frac{\Omega_1(1-\epsilon)}{2\sqrt{2}} & - \Delta
    \end{pmatrix},
\label{eq:hRaman}
\end{equation}
where the rotating-frame basis used is $\{\ket{0},\e{i(\omega_0-\omega_1)t}\ket{-1},\e{i(\omega_0-\omega_1)t}\ket{+1},\e{i\omega_0 t}\ket{e}\}$.

If $\ket{e}$ has a short lifetime, which it often does, it is desirable to minimize the excited state population to avoid spontaneous emission. This can either be done by making $\Delta$ very large compared to all other relevant frequencies, or by using stimulated Raman adiabatic passage (STIRAP). 

\subsection{Raman pulses}

Provided we are interested in dynamics on a timescale that is long compared to the excited state lifetime, we can assume that the excited state amplitude ($a_e$) is damped to equilibrium and adiabatically eliminate the excited state by setting $\dot{a}_e(t) = 0$ in the time-dependent Schr\"odinger equation for Hamiltonian (\ref{eq:hRaman}). Calculating the steady-state of $a_e$ and then substituting back into the equations for the other ground-state amplitudes, we reduce the dynamics to that of a three-level system.  

The effective Hamiltonian is
\begin{equation}
    \mathcal{H}_{\mathrm{eff}} = \hbar \begin{pmatrix}
        -\frac{\Omega_{\rm R}}{2} & \frac{\Omega_{\rm R}(1+\epsilon)}{2\sqrt{2}} & -\frac{\Omega_{\rm R}(1-\epsilon)}{2\sqrt{2}} \\
        \frac{\Omega_{\rm R}(1+\epsilon)}{2\sqrt{2}} & -\delta-\omega_B-\frac{\Omega_{\rm R}(1+\epsilon)^2}{4} & \frac{\Omega_{\rm R}(1-\epsilon^2)}{4} \\
        -\frac{\Omega_{\rm R}(1-\epsilon)}{2\sqrt{2}} & \frac{\Omega_{\rm R}(1-\epsilon^2)}{4} & -\delta+\omega_B-\frac{\Omega_{\rm R}(1-\epsilon)^2}{4}
    \end{pmatrix},
\end{equation}
where we have set $\Omega_0 = \Omega_1 = \Omega$ and made the substitution $\Omega_{\rm R} = \Omega^2/(2\Delta)$, where $\Omega_{\rm R}$ is the effective Rabi frequency. This can be compared to Eq.~(\ref{eq:hrf}) for rf pulses except that now the state energies have acquired ac Stark shifts and the states $\ket{\pm 1}$ are coupled together by a two-photon coupling via $\ket{e}$, with strength $\Omega_{\rm R}(1-\epsilon^2)/4$. The ellipticity parameter $\epsilon$ also has its sign reversed, because the $\sigma^+$ component of the light now addresses the $\ket{-1}$ state rather than the $\ket{+1}$ state.

\begin{figure}[t!]
\centerline{\includegraphics[width=1\columnwidth]{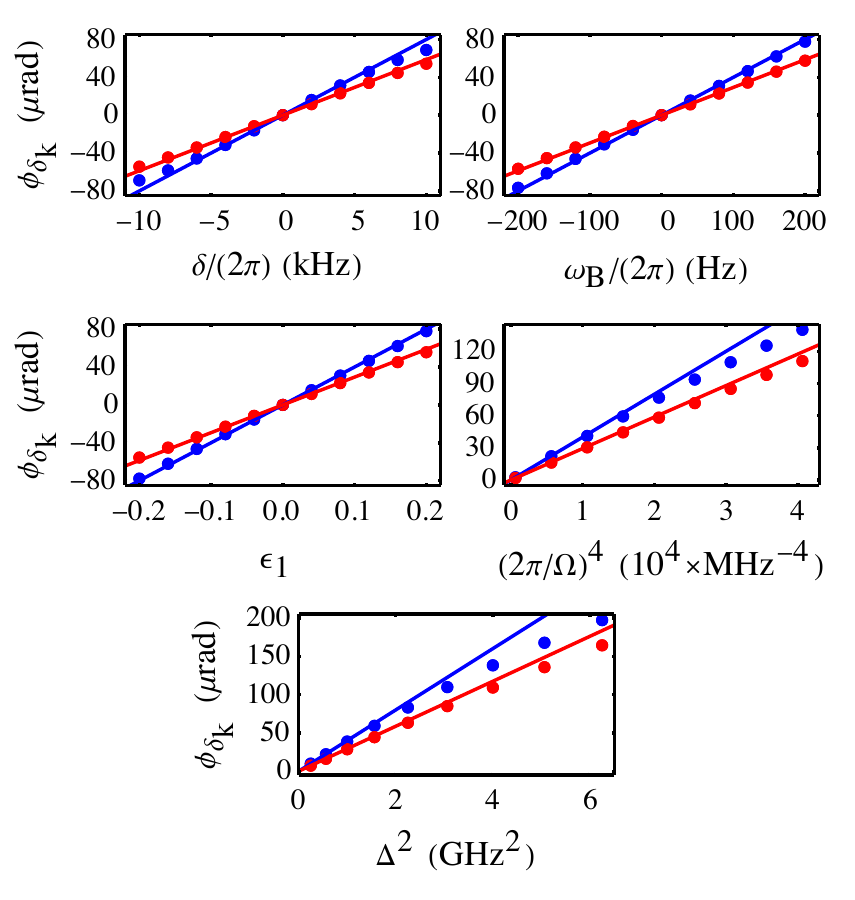}}
\caption{
Comparison of numerical (filled circles) and analytical (lines, Eqs.~(\ref{eq:brffRamanFullInt})) predictions for $\phi_{\delta_k}$, where $\delta_k$ is the two-photon detuning of the optical pulses used for state preparation ($k=1$, in blue) and detection ($k=2$, in red). When not varied, the relevant parameters used are $\Omega = 2\pi\times \SI{10}{MHz}$, $\Delta = 2\pi\times \SI{1}{GHz}$, $\delta = 2\pi \times \SI{5}{kHz}$, $\omega_B = 2\pi\times \SI{100}{Hz}$, $\epsilon_1 = 0.1$ and $\epsilon_2 = 0$.
}
\label{fig5}
\end{figure}

As before, we analytically solve for the asymmetry in the full interferometer using Eq.~(\ref{eq:timeEvolution}) but now substituting $\mathcal{H}_\mathrm{eff}$ for $\mathcal{H}_\mathrm{rf}$. The two optical fields are applied simultaneously for a time $\tau_{\rm R}$ chosen such that $\Omega_{\rm R}\tau_{\rm R} = \pi$, implementing Raman $\pi$-pulses. We carry out the same lowest-order series expansion of the wavefunction in order to obtain an approximate expression for the phase correlated with the two-photon detuning $\delta$. We find
\begin{align}
    \phi_{\delta_1} &= \frac{\omega_B\delta}{\Omega_{\rm R}^2} \left(2\epsilon_1 + \left(\frac{\pi^2}{4}-1\right)\epsilon_2\right) = \frac{4\Delta^2\omega_B\delta}{\Omega^4} \left(2\epsilon_1 + \left(\frac{\pi^2}{4}-1\right)\epsilon_2\right), \nonumber \\
    \phi_{\delta_2} &= \frac{\omega_B\delta}{\Omega_{\rm R}^2} \left(2\epsilon_2 + \left(\frac{\pi^2}{4}-1\right)\epsilon_1\right) = \frac{4\Delta^2\omega_B\delta}{\Omega^4} \left(2\epsilon_2 + \left(\frac{\pi^2}{4}-1\right)\epsilon_1\right),
\label{eq:brffRamanFullInt}
\end{align}
which is similar to Eq.~(\ref{eq:brffFullInt}). The pre-factor for $\epsilon_2$ in the expression for $\phi_{\delta_1}$ is slightly smaller here in the two-photon case compared to the one-photon case and instead of $\Omega$ we have the effective Rabi frequency $\Omega_R$ in the denominator.

We also carry out numerical simulations by solving the time-dependent Schr\"odinger equation for this system. Figure \ref{fig5} compares the results of these simulations to the lowest-order expressions of Eq.~(\ref{eq:brffRamanFullInt}). Unless otherwise stated, we have used a Rabi frequency of $\Omega = 2\pi\times \SI{10}{MHz}$ and a one-photon detuning of $\Delta = 2\pi\times\SI{1}{GHz}$ such that the effective Rabi frequency is $\Omega_R = 2\pi\times\SI{50}{kHz}$. We see good agreement between numerical and analytical results, but with some significant deviations when $\Omega$ is too small or $\Delta$ is too large so that $\Omega_{\rm R}$ becomes comparable with $\delta$. The plots can be compared to those in Fig.~\ref{fig3} where the Rabi frequency was $\Omega = 2\pi\times \SI{100}{kHz}$, showing similar magnitudes of the detuning-correlated phase induced by the imperfect polarization of the light fields.

\subsection{Stimulated Raman adiabatic passage}

\begin{figure}[t!]
\centerline{\includegraphics[width=1\columnwidth]{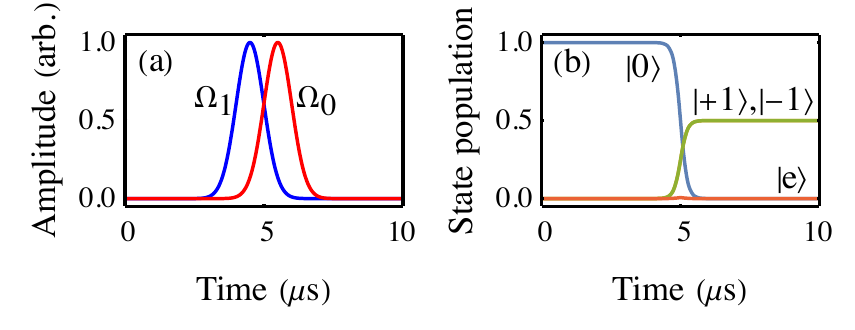}}
\caption{
Stimulated Raman adiabatic passage (STIRAP) with two time-dependent optical fields. 
(a) Pulse sequence showing amplitudes of the two optical fields. (b) State population transfer from $\ket{0}$ to $\ket{x}$, with $\Omega_0 = \Omega_1 = 2\pi\times\SI{10}{MHz}$.
}
\label{fig6}
\end{figure}

Another way to transfer population between $\ket{0}$ and $\ket{x}$ is to use stimulated Raman adiabatic passage (STIRAP). When $\omega_{B}=0$ and $\delta=0$, Hamiltonian (\ref{eq:hRaman}) has two degenerate dark eigenstates,
\begin{align}
    \ket{d_1} &=  \frac{(1-\epsilon)}{\sqrt{2}}\ket{-1} +  \frac{(1+\epsilon)}{\sqrt{2}}\ket{+1}, \nonumber \\
    \ket{d_2} &= \frac{\Omega_1}{\Omega_0}\ket{0} + \frac{(1+\epsilon)}{\sqrt{2}(1+\epsilon^2)}\ket{-1} - \frac{(1-\epsilon)}{\sqrt{2}(1+\epsilon^2)}\ket{+1} \nonumber \\
    &=  \cos\theta\ket{0} + \sin\theta\ket{x''},
\end{align}
where
\begin{equation}
    \ket{x''} = \frac{1}{\sqrt{2}(1+\epsilon^2)}\left( (1+\epsilon)\ket{-1} - (1-\epsilon)\ket{+1} \right) \nonumber
\end{equation}
and $\tan\theta = \Omega_0/\Omega_1$. We have not normalized these states. $\ket{d_2}$  coincides with $\ket{0}$ in the limit where $\Omega_0 \rightarrow 0$, and coincides with $\ket{x''}$ [equivalent to $\ket{x'}$ in Eq.~(\ref{eq:xprimestate})] as $\Omega_1 \rightarrow 0$. Adiabatic evolution between these two limits sweeps $\theta$ from 0 to $\pi/2$, transforming the state from $\ket{0}$ to $\ket{x''}$ while remaining in the dark state at all times. For the simulations presented here, this adiabatic sweep is achieved using the Gaussian pulse sequence shown in Fig.~\ref{fig6}(a). An example of the population transfer is shown in Fig.~\ref{fig6}(b), where we have used a Rabi frequency of $\Omega = 2\pi\times\SI{10}{MHz}$ and set $\Delta = \delta = \omega_B = \epsilon = 0$.

\begin{figure}[t!]
\centerline{\includegraphics[width=1\columnwidth]{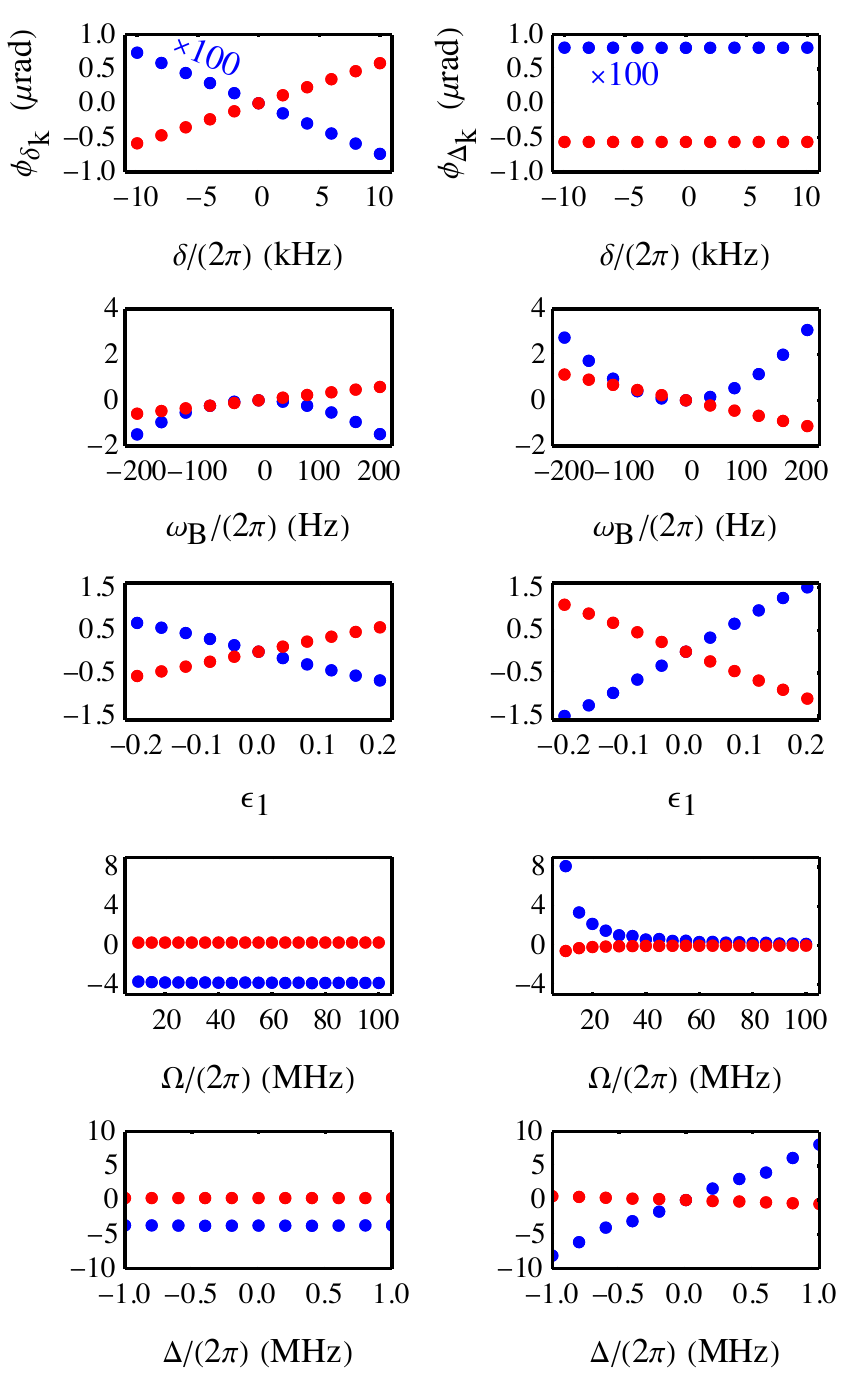}}
\caption{
Numerical simulation results for STIRAP, showing the dependence of the phase on various parameters. The left (right) column shows results for the phase associated with the two-photon (one-photon) detuning $\delta_k$ ($\Delta_k$) of the optical fields used for state preparation ($k=1$, in blue) and detection ($k=2$, in red). For all plots, the points associated with pulse 1 (in blue) are magnified by a factor of 100. The ellipticity of the second set of pulses has been set to zero, $\epsilon_2 = 0$, and the other parameters (when not varied) are $\delta = 2\pi\times\SI{5}{kHz}$, $\Delta = 2\pi\times\SI{1}{MHz}$, $\omega_B = 2\pi\times\SI{100}{Hz}$, $\epsilon_1 = 0.1$ and $\Omega = 2\pi\times\SI{10}{MHz}$. 
}
\label{fig7}
\end{figure}

We are interested in the outcome when there are imperfections. First we note that $\ket{d_2}$ exists irrespective of $\epsilon$ and adiabatic following of the dark state does not depend on $\epsilon$ -- it only influences the state that is reached at the end. Second, we note that the dark states and the adiabatic evolution do not depend on $\Delta$, so we expect the experiment to be highly insensitive to this parameter. Third, we find that when $\delta \ne 0$, $\ket{d_1}$ remains a dark eigenstate (with eigenvalue $-\delta$) but $\ket{d_2}$ is no longer an eigenstate of the system. As $\theta$ slowly increases, the initial state evolves adiabatically towards $\ket{x''}$ but eventually reaches an avoided crossing. If $\delta$ is small enough, this avoided crossing is so small that it will be traversed diabatically, and the target state $\ket{x''}$ will be reached. As $\delta$ increases the avoided crossing opens up and will eventually be traversed adiabatically, returning the system to $\ket{0}$. The same happens when $\Delta$ increases or when $\Omega$ decreases. Here, we consider imperfections that are small enough for the traversal to be strongly diabatic. In this regime, the probability of reaching the final state is robust to the values of the parameters, and we may expect the phases, $\phi_{\Delta}$ and $\phi_{\delta}$ that correlate with the one- and two-photon detunings, $\Delta$ and $\delta$, to be small. We have not found analytical expressions for these phases, so we investigate them by solving the Schr\"{o}dinger equation numerically. The sequence is similar to those studied above -- an initial STIRAP that transfers $\ket{0}$ to $\ket{x''}$, a period of free evolution with a phase $\phi_{B}$ close to $\pm \pi/4$, and then a second STIRAP that in the ideal case, is the reverse of the first. Spontaneous emission has little influence because the excited state population is so small, so is not included in the model.

Figure \ref{fig7} shows results where only the first STIRAP has ellipticity ($\epsilon_2 = 0$). The left column shows how $\phi_{\delta_i}$ depends on the experimental parameters, and the right column shows the same for $\phi_{\Delta_i}$. For each plot, the data for $k=1$ has been magnified by a factor of 100. The first thing we notice is that the phase imperfections from STIRAP are much smaller than those found for the single-photon or Raman processes. The phases correlating with the detunings of the first pulse, $\phi_{\delta_1}$ and $\phi_{\Delta_1}$ are about $10^4$ times smaller, while $\phi_{\delta_2}$ and $\phi_{\Delta_2}$ are about $10^2$ times smaller. Since they dominate, we focus on the latter. We find that $\phi_{\delta_2}$ depends linearly on $\delta$, $\omega_B$ and $\epsilon_1$. It does not depend on $\Delta$ and has very little dependence on $\Omega$ over the range explored here ($\Omega/2\pi$ between 10 and 100~MHz). Similarly, $\phi_{\Delta_2}$ depends linearly on $\Delta$, $\omega_B$ and $\epsilon_1$, but does not depend on $\delta$. It has a $1/\Omega^2$ dependence on the Rabi frequency.

\section{Experimental study}

\begin{figure}[t]
\centerline{\includegraphics[width=1\columnwidth]{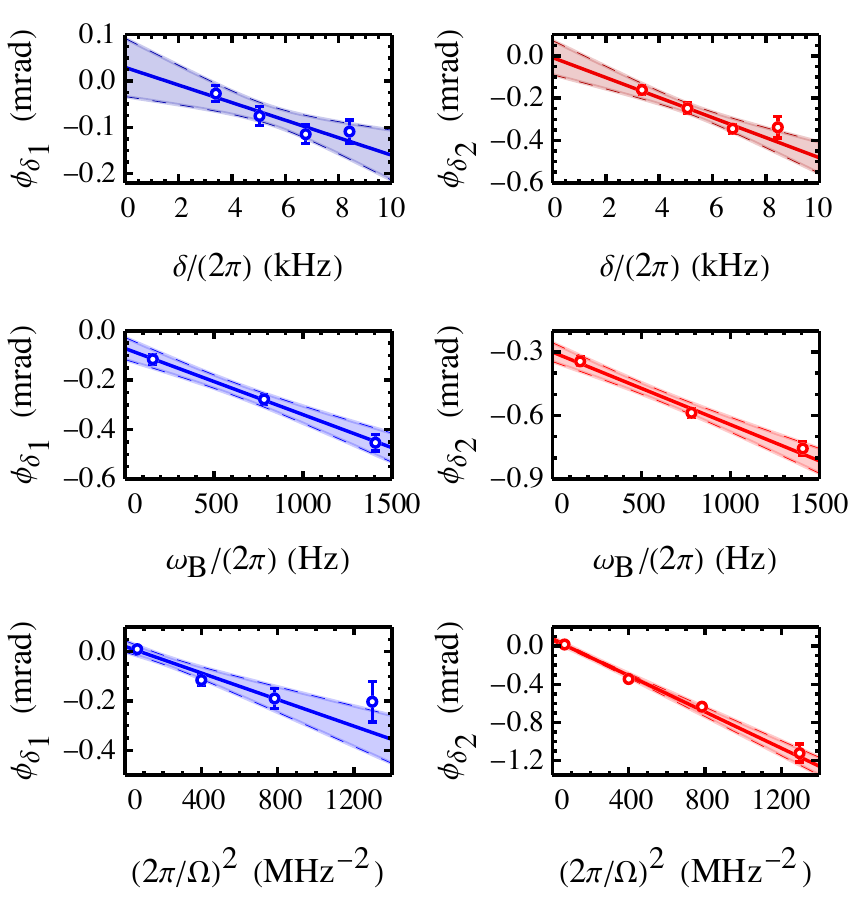}}
\caption{
Experimental measurements of the interferometer phase correlated with rf frequency. The left (right) column shows the variation of the phase correlated with rf pulse 1 (2) with parameters $\delta$, $\omega_B$ and $1/\Omega^2$. When not varied, the parameter values are $\delta/(2\pi) = \SI{6.75}{kHz}$, $\omega_B/(2\pi) = \SI{156}{Hz}$ and $\Omega/(2\pi) = \SI{50}{kHz}$. The solid lines are linear fits to the data, and the dashed lines and shading indicate $2\sigma$ confidence bands.
}
\label{fig8}
\end{figure}

To explore these polarization-dependent systematic effects experimentally, we conducted experiments with single-photon rf pulses using the apparatus illustrated in Fig.~\ref{fig:IFoverview} and described in more detail in \cite{Ho2020}. Although we had no method to control the ellipticities, $\epsilon_{1}$ and $\epsilon_{2}$, we were able to measure their values. To do this, we applied static electric and magnetic fields of $E = \SI{10}{kV/cm}$ and $B = \SI{26}{\mu T}$ along $\mathbf{z}$ so that the $\sigma^\pm$ transitions were separated by several times the rf pulse linewidth. We drove resonant Rabi oscillations by scanning the rf power, and thus determined the power required to drive a $\pi$-pulse on each of the two transitions. The ratio of the powers gives the ellipticity. Fluctuations in the measured ellipticity were significantly larger than the uncertainty of any individual measurement. Accounting for these fluctuations, we estimate $\epsilon_1 = 0.04(2)$ and $\epsilon_2 = -0.003(2)$. Next, with $B$ reduced to its normal (small) value, we collected eEDM data and determined $\phi_{\delta_1(2)}$ from the $\{B\cdot\delta_{1(2)}\}$ channels as described in Section \ref{sec:modelExperiment}.

Figure \ref{fig8} presents our measurements of $\phi_{\delta_{1,2}}$ versus the size of the rf step, $\delta$, the applied magnetic field, expressed as $\omega_{B} = \mu B/\hbar$, and the Rabi frequency, $\Omega$. The magnetic field values were chosen to satisfy $\omega_{B}\tau = n \pi/4$ with $n$ an odd integer. The rf pulse length ($\tau_{\rm rf}$) was varied together with $\Omega$ to satisfy $\Omega \tau_{\rm rf} = \pi$. We see from Figure \ref{fig8} that $\phi_{\delta_{1,2}}$ depends linearly on $\delta$, $\omega_B$ and $1/\Omega^2$, exactly as predicted in Eq.~(\ref{eq:brffFullInt}). Table ~\ref{tab:gradients} gives the gradients of linear fits to these data, and compares them to the gradients expected from Eq.~(\ref{eq:brffFullInt}) and our measurements of $\epsilon_{1,2}$. The gradients with respect to $\omega_B$ agree with the expected values. However, the gradients with respect to $\delta$ and $1/\Omega^2$ are much larger than expected. Furthermore, the y-intercepts of the linear plots against $\omega_B$ in Fig.~\ref{fig8} are non-zero.

\bgroup
\def\arraystretch{1.8}
\setlength\tabcolsep{2pt}
\begin{table}[t!]
\centering
 \begin{tabular}{| c | c | c | c | c | c |}
 \cline{3-6}
 \multicolumn{2}{c|}{} & \multicolumn{2}{c|}{measured} & \multicolumn{2}{c|}{calculated, Eq.~(\ref{eq:brffFullInt})} \\
 \hline
 gradient & units & $i=1$ & $i=2$ & $i=1$ & $i=2$\\
 \hline
$ \frac{\partial \phi_{\delta_i}}{\partial (\delta/2\pi)}$ & $\si{\mu rad/kHz}$  & $-19(6)$ & $-47(7)$ & $-5(3)$ & $-5(3)$ \\
$ \frac{\partial \phi_{\delta_i}}{\partial (\omega_B/2\pi)}$ & $\si{\mu rad/Hz}$ & $-0.27(3)$ & $-0.34(3)$ & $-0.20(11)$ & $-0.23(12)$\\
$\frac{\partial \phi_{\delta_i}}{\partial (2\pi/\Omega)^2}$ & $\si{\mu rad\, MHz^{2}}$ & $-0.27(4)$ & $-0.95(4)$ & $-0.08(4)$ & $-0.09(5)$ \\
 \hline
 \end{tabular}
 \caption{Measured and calculated gradients of the phases $\phi_{\delta_i}$ with respect to $\delta$, $\omega_B$ and $1/\Omega^2$. The measured gradients, with their $1\sigma$ uncertainties, are extracted from the linear fits in Fig.~\ref{fig8}, while the calculated gradients use the known parameter values and measured rf ellipticities.}
 \label{tab:gradients}
\end{table}
\egroup

These observations are all consistent with the hypothesis of two separate contributions to $\phi_{\delta}$ -- the one given by Eq.~(\ref{eq:brffFullInt}) plus a second effect that is proportional to $\delta/\Omega^2$ but independent of $\omega_{B}$. This second effect could potentially be caused by the combination of an offset detuning ($\Delta_i$) and an interferometer phase that does not reverse with $\switch{B}$ ($\phi_{\rm bg}$), as can be seen from Eq.~(\ref{eq:offsetBdelta}). However, this effect has already been eliminated through measurement of the $\channel{B}$ and $\channel{\delta_i}$ channels, as described in Sec.~\ref{sec:param_imperfections}. We have to conclude that either our method of eliminating the effect described by Eq.~(\ref{eq:offsetBdelta}) is inaccurate, or that there is yet another contribution to $\phi_{\delta}$ that is not revealed by the analysis presented in this paper.

\section{Conclusions}

Electron EDM measurements use radiation to prepare and read out the spin state. We have studied how ellipticity in the polarization of this radiation leads to an interferometer phase that correlates with the detuning from resonance. An imperfect electric field reversal changes the detuning via the Stark shift. Together, these two effects produce an $E$-correlated phase that is a systematic error in the eEDM measurement. This systematic error is linear in the ellipticity, the Zeeman splitting and the $E$-correlated detuning, and scales inversely as the square of the Rabi frequency. The phase correlating with detuning and the change in detuning correlating with $E$ are imperfections that are typically measured automatically in an eEDM measurement. Those measurements can be used to minimize the imperfections, and to correct the residual systematic error when imperfections remain. We have found approximate analytical expressions for the systematic error and its correction in the case where the state preparation and readout use a coherent single-photon process or a Raman process. Numerical simulations confirm the accuracy of these expressions. 

We have also studied this type of systematic error experimentally, using rf pulses for state preparation and readout. We find evidence for an interferometer phase that correlates with rf detuning and is consistent with the ellipticity-induced effect. However, there is a second effect that produces a phase correlated with detuning -- it is distinguishable from the ellipticity effect since it does not depend on the Zeeman splitting. We have not yet established the source of this second effect.

When STIRAP is used for state preparation and readout, the systematic error is found numerically to be about two orders of magnitude smaller. Thus, STIRAP not only offers the benefits of a robust, all-optical approach to state manipulation, but also provides relative immunity to an important class of systematic error.

\begin{acknowledgments}
 
This work was supported by funding in part from the
Science and Technology Facilities Council (grants ST/S000011/1 and ST/V00428X/1); the Engineering and Physical Sciences Research council (grant EP/X030180/1); the Sloan Foundation (grant G-2019-12505); and the Gordon and Betty Moore Foundation (grant 8864). The opinions expressed in this publication are those of the
authors and do not necessarily reflect the views of these funding bodies.

 \end{acknowledgments}

\end{document}